\documentclass{optica-article}
\journal{opticajournal} 
\articletype{Research Article}

\usepackage{subcaption}
\usepackage{gensymb}

\usepackage{wrapfig}
\usepackage{sidecap}
\begin{document}
\title{A compact, dual-colour chirped-pulse laser system for field-deployable coherent Rayleigh-Brillouin scattering diagnostics
}

\author{Yingjie Zhao,\authormark{1,2} Gabriel M. Flores Alfaro,\authormark{1,2}, Carmen Pascual-Fort,\authormark{3}, Gabi-Daniel Stancu,\authormark{3} and Alexandros Gerakis\authormark{1,4,*}}

\address{\authormark{1}Luxembourg Institute of Science \& Technology, Belvaux L-4362, Luxembourg\\
\authormark{2}Université du Luxembourg, 2 avenue de l’université, Esch-sur-Alzette L-4365, Luxembourg\\
\authormark{3}Laboratoire EM2C, CNRS, CentraleSup\'{e}lec, Universit\'{e} Paris-Saclay,
3 rue Joliot Curie, 91192, Gif-sur-Yvette Cedex, France\\
\authormark{4}Department of Aerospace Engineering, Texas A\&M University, College Station, Texas 77845, USA}

\email{\authormark{*}alexandros.gerakis@list.lu} 





\begin{abstract}

Single-shot coherent Rayleigh-Brillouin scattering (CRBS) requires dual colour, frequency, and pulse duration agile laser systems, currently available only in laboratory layouts. In this work, we present a compact, portable, and robust next generation laser system that meets all of these requirements. Key features include an integrated master-slave seed laser design for stable frequency chirping, fiber-coupled acousto-optic modulator (AOM) gated pulse shaping for long-term thermal stability, and customized amplification stages in a rugged, portable enclosure which simplifiy the alignment and maintenance process. 
The system in this implementation can deliver up to $450$~mJ per arm, chirp rates up to $\sim22$~MHz$/$ns, and arbitrary temporal profiles within $\sim25$~ns to $\sim1$~$\mu$s. This laser system is validated by measuring the neutral number density of an atmospheric plasma jet, and a two-dimensional temperature map is constructed, demonstrating lab-grade diagnostics for field deployment.
\end{abstract}



\section{Introduction}
Coherent Rayleigh-Brillouin scattering (CRBS) is a seedless, four-wave mixing, non-intrusive laser diagnostic technique for probing thermodynamic properties such as number density, translational temperature and flow velocity \cite{grinstead2000coherent,pan2002coherent,Gerakis2016}. Its capability for neutral gas flow velocimetry, reactive combustion flows and plasma environments has been investigated~\cite{gerakis2025seedless, Bookey:06, randolph2026density}. Additionally, techniques based on the same underlying principle have been proposed for charged-particle diagnostics in high-pressure, low-temperature plasmas~\cite{10.1063/5.0072540, FloresAlfaro2024Langmuir,FloresAlfaro2025BrillouinPlasmas}.

In CRBS, two counter-propagating laser beams, called \textit{pumps}, form an interference pattern where polarizable molecules or atoms are attracted to the high intensity regions of the resulting interference pattern by the dipole force and subsequently create an optical lattice. A third beam, called the \textit{probe}, is incident at the Bragg angle to this optical lattice, and forms a fourth coherent laser beam, termed the \textit{signal} \cite{pan2002coherent,grinstead2000coherent}.
In its earlier implementations\cite{grinstead2000coherent,pan2003coherent,PhysRevLett.107.173903}, CRBS required the use of broadband pump beams to simultaneously generate optical lattices with different phase velocities, while the power spectrum of the signal (i.e. the CRBS lineshape) was resolved through an etalon with frequency scanning of the probe beam. The CRBS spectral acquisition was thus limited by the etalon and/or the probe laser scanning speeds, resulting in spectral acquisition times on the order of 10's of minutes; this placed significant practical limits in the use of CRBS in highly transient environments. To overcome this limitation, single-shot CRBS was developed, in which a chirped pump beam is introduced to generate an optical lattice with variable phase velocity during the duration of the laser pulse~\cite{gerakis2013single}. This enabled the velocity distribution function (VDF) of the probed species to be mapped in a single laser shot of $\sim150-300$~ns duration. Currently, this technique has been expanded from a single point to a simultaneous multipoint measurement, allowing e.g. gas characterization of boundary layers above airfoils in a wind tunnel~\cite{doi:10.2514/6.2025-0242, Kumar2025Multipoint}.

Although single-shot CRBS offers unique advantages such as high signal-to-noise ratio (SNR) and the fact that it is non-resonant~\cite{bak2023torr}, to this day, it has not been as widely adopted by the relevant communities, as it would have been anticipated. This is largely attributed to the complexity of the necessary laser instrumentation required for successful implementation of this technique. Commercial lasers do not offer a combination of pulse shape agility (from $\sim25$ ns to $\sim1$ $\mu$s), chirp rate (GHz bandwidth), and pulse energy ($\geq250$~mJ). Previous work has demonstrated nanosecond pulse shaping using electro-optical modulation for pulse shaping and chirping, as well as high-energy Nd:YAG amplification with arbitrary temporal control \cite{Meijer:17,Rogers:16,Malinowski:09,beyer2021frequency}. However, integrating these capabilities with stable chirped operation suitable for single-shot CRBS diagnostics remains non-trivial and often requires large laboratory-scale installations. For example, Bak et al. demonstrated a custom dual-colour, single-shot CRBS laser system occupying a $3\times1.2\times1.6$~m optical table, weighing in total in excess of $1500$~Kg, which limits experimentation to environments and apparatus developed in close vicinity (see e.g. Ref. \citenum{Bak2022DualAgile}). This system has also recently been upgraded to deliver pulse energy of $\sim2$~J on the same laboratory scale~\cite{11298376}. While these systems exhibit relatively stable and reliable operation and their use has been demonstrated in a laboratory environment~\cite{Kumar2025Multipoint}, the demand for real-world scientific and engineering scenarios, such as full-scale wind tunnels, jet engines, matter transport and plasma jets~\cite{dogariu2019single, osti_3001153}, requires a portable, robust, and easily maintainable next-generation single-shot CRBS laser system for field deployment.

Here, we present a new generation single-shot CRBS laser system which maintains the important characteristics of the previously developed laser systems  while being portable, compact and robust; importantly, the main laser head now only measures $1\times1.2$~m while weighing $\sim80$~kg. The main architecture of this laser system is based on the successful implementation presented in Refs.~\citenum{10.1117/12.2253421, Bak2022DualAgile,11298376}; however, the new system delivers increased performance and stability through the use of a novel pulse shaping scheme, within a significantly smaller and rugged enclosure. Its capability is validated through in-situ CRBS measurements of an atmospheric plasma jet, demonstrating that laboratory-grade CRBS diagnostics can be deployed directly at the point or facility of need.

\section{\label{sec:level2}CRBS laser design}

\begin{figure*}[h!]
    \centering
\includegraphics[width=0.99\textwidth]{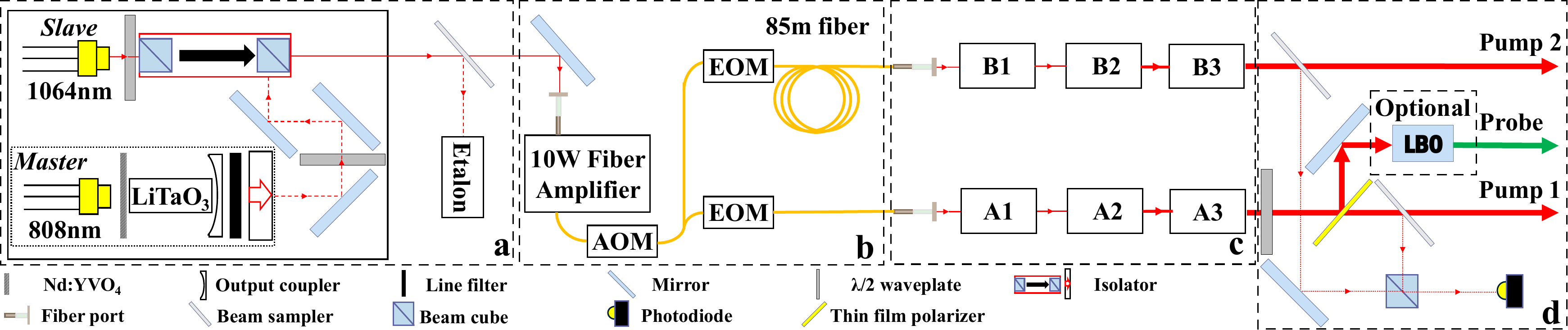}
\caption{\label{fig:CRBSscheme}The scheme of the CRBS laser system with 4 sections: a) \textit{Master-slave} injection locking. b) Pulse shaping. c) Amplification in tandem. d) Beam output. A1/B1, A2/B2 and A3/B3 illustrate the amplifier configuration and beam path. In section a), \textit{master} laser is within the dashed-line rectangle, and the output of \textit{slave} diode laser is sampled by a pick-off glass into an etalon to monitor the injection locking. In section b), a $85$~m fiber is used to achieve the temporal alignment of the chirped and unchirped beam. In section c), optics for increasing the beam diameter between each stages are not shown. In section d), a LBO crystal is optional for frequency-doubled probe beam, and each pump beam is sampled for heterodyne detection.}
\end{figure*}

The portable single-shot CRBS laser system presented here consists of $4$ sections, respectively presented in Fig.~\ref{fig:CRBSscheme}: a) Master-slave injection locking: 
b) CW pre-amplification and pulse shaping
c) Amplification in tandem and d) Beam output. In the following sections, we present in detail the operation of the different sections. 


\subsection{\label{sec:level2-1}Master-slave injection locking}



For single-shot CRBS, one fixed frequency pump beam and one chirped pump beam are required in order to generate an optical lattice with a continuously varying phase velocity in a single laser pulse. This chirping imposes strong demands on the laser system: the frequency must sweep several GHz to cover the required phase velocity space, while the pulse energy and temporal lineshape must remain stable both shot-to-shot and during the duration of the pulse. To achieve fast frequency chirping while maintaining stable output power, a \textit{master–slave} architecture is adopted, as illustrated in Fig.~\ref{fig:CRBSscheme}(a). The frequency chirp is achieved by a custom, in-house designed and built diode-pumped solid-state (DPSS) laser, referred to as the \textit{master} laser, whose output is used to injection-lock a CW diode laser, referred to as the \textit{slave} laser. The $808$~nm pump laser diode, the cavity with gain medium, and the slave laser diode are mounted on individual temperature-controlled copper blocks, which are all fixed on a bigger copper baseplate serving as a thermal sink. This integrated design yields a more compact and thermally stable configuration compared to the previously reported separated layout~\cite{Bak2022DualAgile}.


The \textit{master} laser uses a Nd:YVO\textsubscript{4} crystal as the gain medium\footnote{$2.95\times3.05\times0.5$~mm with $3$~\% Nd doped; CASIX} which is pumped by an $808$~nm diode laser\footnote{L808P1000MM; Thorlabs}. A $100$~mm ROC plano-concave output coupler\footnote{101585; Layertec} with $97.5$~\% reflectivity is chosen such that the output power is maximal. A LiTaO\textsubscript{3} Electro-Optic modulator (EOM) crystal\footnote{$1\times3\times5$ mm; CASIX} is placed inside the cavity for frequency chirping: by applying a voltage to the crystal, its refractive index changes accordingly, resulting in a variation of the optical cavity length and thus modulation of the output laser frequency. This frequency modulation $\Delta f$ with respect to the voltage $V$ applied to the crystal is given by~\cite{10.1063/1.1499541}:

\[
\Delta f = \frac{f\, n_e^{3}\, l\, r_{33}}{2\, l_{\mathrm{cav}}\, d}\, V= R_{\text{EO}} V,
\]
where $f$ is the fundamental optical frequency of the cavity, $n_e$, $l$ and $d$ are the refractive index, length and thickness of the EOM crystal, respectively, $l_{\mathrm{cav}}$ is the cavity length and $R_{\text{EO}}$ is the effective modulation coefficient.
The chirp rate $c=\Delta f/ \Delta t$ can be expressed as $c=R_{\text{EO}}V/\Delta t$. For a fixed voltage amplitude and modulation coefficient, increasing the chirp rate demands a higher frequency for the applied voltage waveform. A function generator~\footnote{3314A; HP} provides a ramp signal that is then amplified by a voltage amplifier~\footnote{A-304; A.A. Lab Systems}. Due to the limited bandwidth of the amplifier, the maximum practical frequency of this driving waveform is $800$~kHz. Beyond this point, further increases in frequency no longer contribute to a higher chirp rate~\cite{Bak2022DualAgile}.

While placing the EOM crystal inside the cavity provides efficient frequency chirping, it simultaneously introduces several causes that substantially render the output power unstable. One major issue is relaxation oscillations~\cite{Koechner_SolidStateLaserEngineering}: these occur since the upper-state lifetime of Nd:YVO\textsubscript{4} ($\sim90$~$\mu$s~\cite{GUILLOTNOEL2000427}) is orders of magnitude longer than the cavity round-trip time ($\sim33$~ps for a $5$~mm cavity) and the gain medium that cannot follow rapid intracavity perturbations on a round-trip timescale. Consequently, when a voltage waveform is applied to the intracavity EOM, it perturbs the laser from its steady state by perturbing the balance between stimulated emission and population inversion, and thus the resulting output power oscillates. Since the Nd:YVO\textsubscript{4} gain medium is strongly birefringent and oscillates along a fixed polarization axis, these polarization fluctuations also lead to time-dependent intracavity losses. As a result, the output power becomes unstable and further relaxation oscillations can be induced. 

Due to the relaxation oscillations caused by the intra-cavity EO modulation, the output power of the \textit{master} laser fluctuates after the voltage is applied~\cite{Bak2022DualAgile}. Transmitting this unstable laser beam directly to the pulse shaping and amplification blocs would distort the temporal lineshape of the resulting final pulses and those oscillations would eventually get amplified as well, resulting in an unstable CRBS signal - the intensity of which is dependent on the square of the intensities of the pump and probe beams. 

To decouple the frequency information from the power instability, the output of the \textit{master} laser is injected into a $1064$~nm CW single-mode diode laser~\footnote{M9-A64-0200; Sheaumann}, which serves as the \textit{slave} laser. Through optical injection locking, a fraction of the master light enters the slave cavity through its output. When the instantaneous frequency of the injected beam lies within the locking range of the slave resonator, the injected field suppresses the \textit{slave}'s free-running mode and forces it to oscillate exactly at the \textit{master}'s instantaneous frequency~\cite{8859279}.

In the end, the resulting \textit{Master-Slave} laser delivers a CW beam that combines both power stability (determined by the \textit{slave} diode steady-stage operation) and agile frequency chirping (imposed by EO modulation of the \textit{master} laser). The chirp rate achieved by this combined system, measured after subsequent pulse amplification, is characterized in Section~\ref{sec:level2-4}.

\subsection{\label{sec:level2-2}CW pre-amplification and pulse shaping}
Since the CRBS signal is proportional to the pump and probe beam intensities, a higher achievable SNR requires a stronger intensity. This can be achieved by increasing the pulse energy and/or decreasing the pulse duration. Hence, a pulsed beam is preferred over a CW beam, as it can give a much higher peak power by orders of magnitude compared with the average power of a CW laser beam. Here, this pulse shaping capability is achieved by intensity EOMs (iEOMs)~\footnote{AM1064b; Jenoptik} controlled by an arbitrary function generator (AFG)~\footnote{AFG3102C; Tektronix}. Also, single-shot CRBS requires two \textit{pump} beams that are temporally aligned to create the optical lattice, yet both generated from the same CW \textit{Master-slave} laser source: when the drive voltage is active to the EOM crystal, the output is chirped and when no voltage is applied, the output is at the fundamental optical frequency. Since these two outputs originate from the same optical cavity, they are self-referenced in phase and frequency, which is a significant advantage over seeding the two beams from independent laser sources~\cite{cornella2013narrowband}.

To increase the overall intensity, the CW output of the \textit{Master-Slave} laser is initially injected in a commercial $10$~W Yb-based fiber amplifier~\footnote{YAR-10K-1064-LP-SF; IPG Photonics}, before being split into 2 arms by a $50:50$ fiber coupler. However, to ensure the stability of the system and the safety of optics, especially protecting the iEOM under its maximum input power ($300$~mW), the output power of the fiber amplifier does not exceed $1$~W. This maximum power is set by gradually increasing the output power of the fiber amplifier while measuring the input power of both EOMs until they reach $300$~mW. As seen in Fig.~\ref{fig:CRBSscheme}b), each iEOM receives both chirped and unchirped CW beam components, thus pulse picking is essential to separate the two outputs: the chirped pulse is extracted from one iEOM, while the unchirped pulse is extracted from the other. 

Typically, the CRBS laser pulse has a width of $250$~ns\cite{gerakis2013single} at a repetition rate of $10$~Hz, meaning that the iEOMs let through only up to $0.0025$~\% of the input energy, with the remaining energy being dissipated as heat in the iEOM. This excessive heat causes the optimal voltage level of the iEOMs to drift, leading to distortions in the temporal profile of the pulse, while another well-known cause for this unstable phase shift is pyroelectric effect~\cite{Yamada_1981,salvestrini2011analysis}. In practice, this drift could even be observed in several minutes after the voltage level to the iEOMs is set.

To overcome this issue, a fiber-coupled acousto-optic modulator (AOM)~\footnote{1064-AM-1; Aerodiode} is added between the fiber amplifier and iEOMs working as an optical gate to preshape the CW beam into a longer pulse. The AOM is controlled by an RF driver~\footnote{RFAOM-T-100; Aerodiode} with TTL digital input from the delay generator~\footnote{DG645; Stanford Research System}. When the AOM is driven by the RF driver, an acoustic wave is launched within its crystal, causing the input CW beam to be Bragg scattered from this wave thus forming the so-called first-order output which acts as input to the iEOMs. When the driver is off, there is no beam output from the AOM, thus a laser pulse can be generated by triggering the driver as off-on-off. One should notice that the rise and fall time of the AOM output is much longer than that of the iEOM ($50$~ns compared to $200$~ps). In this case, the duration of the AOM output should be set at least 2 times longer than the target duration of the pulse, ensuring the iEOM can receive both chirped and unchirped CW input.

By adding the AOM for pulse preshaping in the system, the duty cycle of the iEOM is greatly increased from $0.0025$\% to $33.3$\% with a $750$~ns pulse of AOM preshaping. This improvement significantly reduces heat dissipation on the iEOM and, as a result, the optimal voltage level of the iEOM is stable for substantially longer times on the order of several hours. In addition, this improves the overall performance of the resulting CRBS portable system, such as the output power and the quality of the pulse shape~\cite{Meijer:17}. This optimization ensures the robustness of the system when working for a long period at high power.

\begin{SCfigure}[1.0]
    \centering
    \includegraphics[scale=0.4]{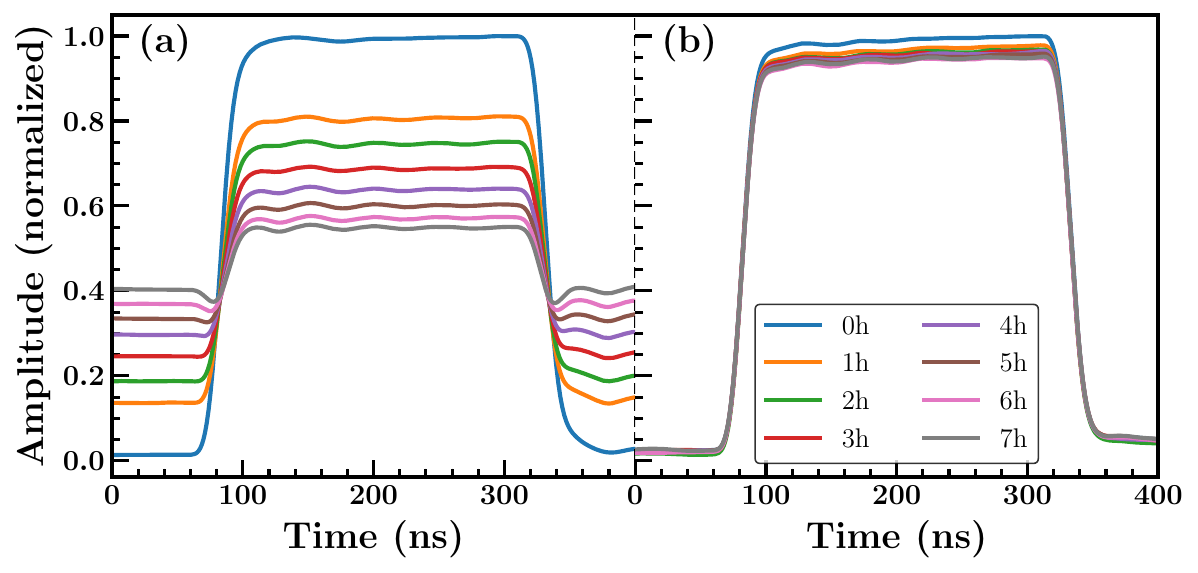}
    \caption{\label{AOMEOM}
    Long-term square pulse stability comparison of EOM output: (a) without AOM, where CW leakage and thermal effects cause progressive pulse deterioration, and (b) with an upstream AOM, which suppresses leakage and stabilizes the thermal load, maintaining pulse shape and amplitude. Measurements were recorded over a 7-hour period at 1-hour intervals.}
\end{SCfigure}

Here, we compare the pulse shape stability of two iEOM configurations over a 7-hour period: one iEOM is paired with an upstream AOM, while the other modulates directly the fiber amplifier output. As shown in Fig.~\ref{AOMEOM}(a), without the AOM, the top of the square pulse gradually droops and the baseline rises over time, indicating the drift of optimal voltage level caused by the thermal effects. Fig.~\ref{AOMEOM}(b) shows that the upstream AOM maintains a stable pulse shape, as the AOM acts as an optical gate that blocks the CW leakage between pulses and stabilizes the thermal load on the iEOM.

Since the chirped and unchirped pulses are picked at different iEOM gating times, as described above, an $85$~m fiber is introduced as a delay line to compensate for the resulting temporal offset. Fine temporal alignment between two pulses is achieved by adjusting the delay between each channel of the AFG while monitoring the heterodyne signal.

\subsection{\label{sec:level2-3}Amplification in tandem}
After pulse shaping, the pulses are coupled into free-space amplification chains via fiber ports. Each amplification arm is equipped with four diode-pumped Nd:YAG amplifiers arranged in a double-pass configuration, including one RBAT20, two RBAT34, and one REA7006 modules~\footnote{Northrop Grumman Cutting Edge Optics}, as illustrated in Fig.~\ref{fig:CRBSscheme}c. A Pockels Cell (PC) is positioned after the first amplifier and operates at quarter voltage ($\sim$3.4~kV), functioning as a quarter wave plate. During pulse transmission, the quarter voltage is applied to rotate the polarization such that the amplified pulse is transmitted while suppressing parasitic lasing and preventing optical feedback. In contrast to previous implementations~\cite{Bak2022DualAgile}, where the PC operated at half voltage ($\sim$6.8~kV), this configuration reduces the electrical stress on the crystal and improves the PC lifetime. In the second amplification stage A2/B2, two identical amplifiers are placed in tandem. Compared to a previous laboratory system where each RBAT34 amplifier has a separate stage~\cite{Bak2022DualAgile,11298376}, this configuration reduces the overall size and the number of required optical components, thus allowing a more compact and lower cost design. However, combining two amplifiers within a single stage can introduce depolarization losses, which may reduce the output energy and degrade the beam profile \cite{Koechner_SolidStateLaserEngineering}. To address this issue, a~$90\degree$ quartz rotator~\footnote{470-4649,; EKSMA} is inserted between the two amplifiers. This approach suppresses depolarization effects, improves transmission efficiency, and consequently enhances the output energy and overall performance of the portable CRBS laser system.

\begin{figure}[h]
    \centering
\includegraphics[scale=0.3]{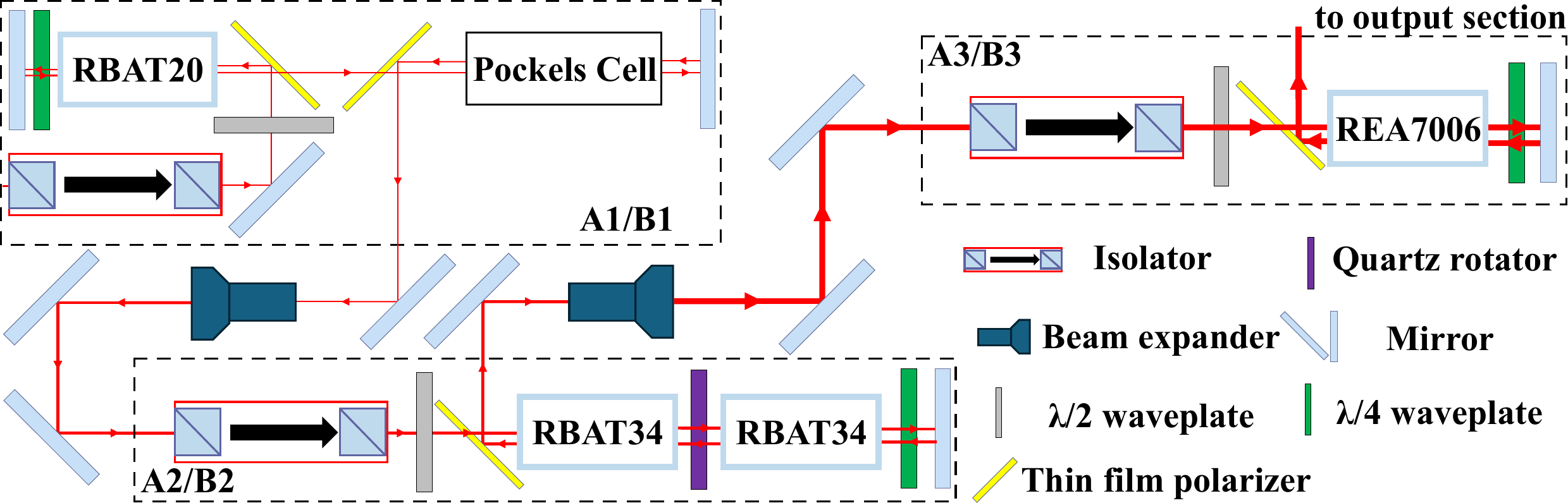}
\caption{\label{baseplate}Schematic of the free-space amplification chain 
(only one arm is shown). The beam enters A1/B1 stage via an isolator, then double-passes through a RBAT20 amplifier, then followed by a Pockels Cell operating at quarter-wave voltage, which suppress parasitic lasing and optical feedback. In the second stage A2/B2, two RBAT34 amplifiers are arranged in tandem with a~$90\degree$ quartz rotator to compensate depolarization loss. The REA7006 amplifier at stage A3/B3 brings the pulse to full energy before being delivered to the output section. Component positions are schematic and do not represent the true scale or spacing of the optical elements.}
\end{figure}

These amplifiers, along with other optical components, are mounted on a custom-designed $20$~mm thick base plate, ensuring a consistent beam height
, as shown in Fig~\ref{baseplate}. 
The positioning of all optical components is performed by aligning optical centers within the CAD assembly; prior to machining, a 3D printing base plate prototype was used to verify if the alignment of the design is achievable. 

Some optics such as mirrors and thin-film polarizers (TFPs), are mounted on commercial kinematic mounts that do not necessarily have the same height from their optical axes to the contact surface. In the initial design, a base plate $\sim$40 mm thick was selected, and the mounts were recessed into the base plate by removing material to fulfill their shape. Although this configuration was quite stable, its weight exceeded 100 kg, making transport difficult. Furthermore, some optomechanical parts are fastened from below, making maintenance challenging when any optical component or mount needed to be replaced.

To solve the overweight and maintainability issues, a $20$~mm thickness base plate was designed. This thinner design significantly reduces the weight of amplification stages, and the optical component fasteners are accessible from above. To keep the optical axis height at $63.5$~mm, custom-made rising holders are developed for each optical and optomechanical parts. This new design simplifies the maintenance process while also providing the option to easily upgrade optics and mounts.

\subsection{\label{sec:level2-4}Beam output}

\begin{figure}[h!]
\centering

\begin{subfigure}[b]{0.49\linewidth}
    \centering
    \includegraphics[width=\linewidth]{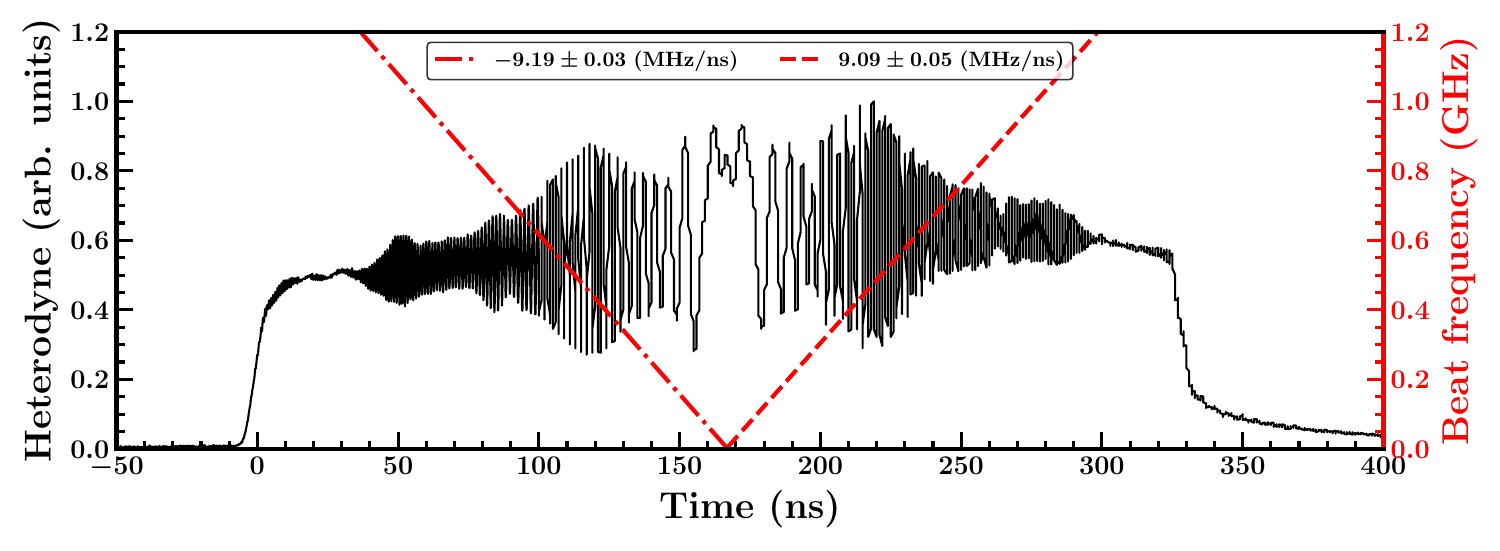}
    \caption{Heterodyne measured with a $1$~GHz bandwidth photodiode. On the edge of the pulse, the instantaneous frequency is not resolved due to insufficient bandwidth. The chirp rates are  $-9.19 \pm 0.03$~MHz/ns and $9.09 \pm 0.05$~MHz/ns respectively, with $120$~V applied to the EOM crystal.}
    \label{fig:hte}
\end{subfigure}
\hfill
\begin{subfigure}[b]{0.49\linewidth}
    \centering
    \includegraphics[width=\linewidth]{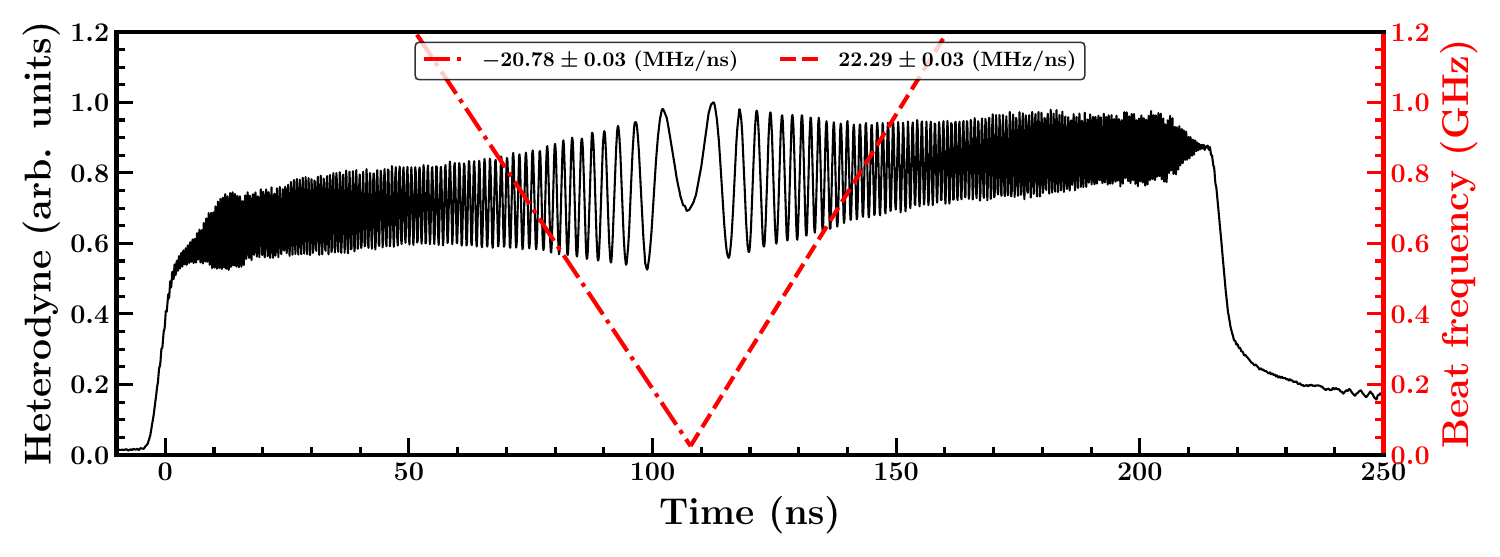}
    \caption{Heterodyne measured with a $5$~GHz photodiode. The chirp rates are $-20.78 \pm 0.03$~MHz/ns and $22.29 \pm 0.03$~MHz/ns respectively, with $244$~V applied to the EOM crystal. We mention that the observed pedestal is attributed to the longer fall time of the $5$~GHz photodiode used.}
    \label{fig:chirp1}
\end{subfigure}

\caption{\label{chirp}
Heterodyne measurements of the amplified pulses. The black solid curve shows the beat signal, obtained by mixing the chirped arm with the unchirped arm. The red dot-dashed and dashed lines are linear fits to the instantaneous frequency.}
\end{figure}

After amplification, each arm delivers pulses with energies up to $450$~mJ/pulse. The achievable chirp of the amplified pulse is evaluated by heterodyne detection, in which the beam output of the chirped arm is heterodyned with the beam of the unchirped arm on a $5$~GHz photodiode~\footnote{DET08C; Thorlabs}. Through analysis of the heterodyne signal, it is estimated that the chirp rate can reach $22.29 \pm 0.03$~MHz/ns with up to $244$~V applied to the EOM crystal. This is about $20\%$ higher than the system in Refs.\citenum{Bak2022DualAgile,11298376} with the same voltage applied. As shown in Fig.~\ref{chirp}, two asymmetric linear chirp rates are produced since the amplified ramp voltage signal is distorted~\cite{11298376}. By tuning the relative delay between channels on the AFG, the two beams can be precisely synchronized in time.

As shown in Fig.~\ref{fig:CRBSscheme}d), for single-shot CRBS one arm serves as a pump beam, while the other is further split into pump and probe beams using a half-wave plate and a TFP, with the splitting ratio controlled by rotating the half-wave plate.  The probe beam can be frequency doubled by an LBO crystal~\footnote{$4\times4\times50$~mm; Castech} housed in a compact, temperature-stabilized oven~\footnote{TC2 and CO1; EKSMA}. This dual-colour configuration is adopted to enhance the signal-to-noise ratio at low pressures: the shorter probe wavelength yields a sixteen-fold increase in scattered signal intensity through the $\lambda^{-4}$ dependence of Rayleigh–Brillouin scattering, while also enabling the use of high-sensitivity PMT detectors whose quantum efficiency is substantially higher at $532$~nm than at $1064$~nm~\cite{bak2023torr}. 
\newpage

\begin{wrapfigure}{r}{0.5\textwidth}
    \centering
    \includegraphics[width=0.5\textwidth]{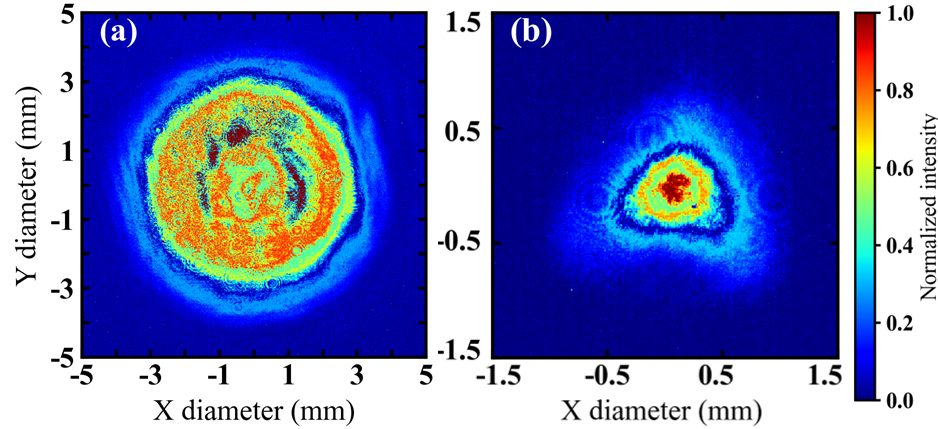}
    \caption{\label{Beamprofil}
    Normalized near-field beam profiles of (a) the $1064$~nm probe beam with around $6$~mm diameter, measured at the reflection of a wedge, exhibiting ring-like structures associated with higher-order transverse modes from the Nd:YAG rod amplifier, and (b) the corresponding $532$~nm second harmonic beam, measured directly with a beam profiler, showing a more spatially confined profile due to the intensity-dependent nature of the SHG process.}
\end{wrapfigure}

\begin{figure}[htbp]
    \centering

    \begin{subfigure}{0.49\textwidth}
        \centering
        \includegraphics[width=\linewidth]{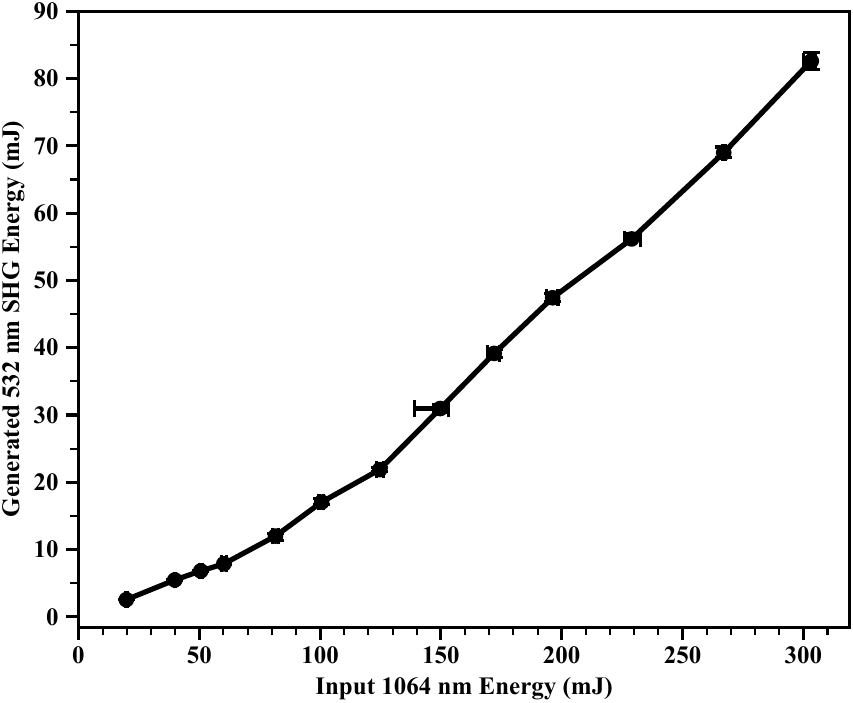}
        \caption{\label{BaseSHG}
        Generated $532$~nm beam energy as a function of input $1064$~nm pulse energy, showing a near-quadratic dependence at low energy regime. Each data point is averaged over 40 pulses, and the error bars represent the standard deviation of the energy measured by the energy meter.}
    \end{subfigure}
    \hfill
    \begin{subfigure}{0.49\textwidth}
        \centering
        \includegraphics[scale=0.4]{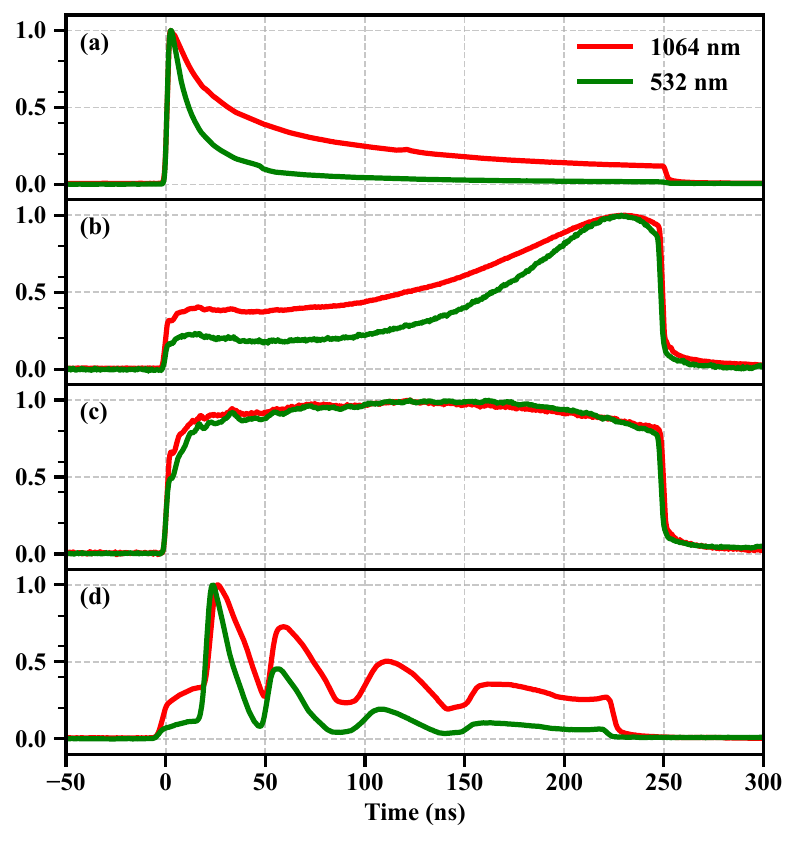}
        \caption{\label{SHG}
        Normalized pulse shapes of the $1064$~nm input (red) and the corresponding $532$~nm SHG output (green) for four distinct temporal profiles programmed via the EOM and AFG: (a) exponentially decaying, (b) ramp-up, (c) multi-peak, and (d) flat-top.}
    \end{subfigure}
\caption{\label{532energypulseshape}
Second harmonic generation conversion efficiency and normalized temporal pulse shapes of the  $1064$~nmg input and generated $532$~nm beams.}
\end{figure}

The near-field beam profile of the $1064$~nm probe beam, measured at the reflection of a wedge, is shown in Fig.~\ref{Beamprofil}(a). The profile has ring-like structures, showing that higher-order transverse modes are obtained, which is characteristic of large-aperture Nd:YAG rod amplifiers operating at high gain. The generated $532$~nm beam profile is shown in Fig.~\ref{Beamprofil}(b), which is more spatially confined than the fundamental, as the intensity-dependent nature of the SHG process preferentially converts the higher-intensity central portion of the beam. The SHG conversion efficiency reaches approximately $17$~\% at a typical probe beam input energy of $100$~mJ, as shown in Fig.~\ref{532energypulseshape}(a), and the conversion efficiency is  proportional to the input energy, which is consistent with the expected behavior of second-order nonlinear conversion in the low-depletion regime.

To demonstrate the temporal shaping capability of the system, four different pulses programmed via the iEOM and AFG are used as input to the LBO crystal, with the corresponding $532$~nm outputs shown in Fig.~\ref{532energypulseshape}(b). The SHG output follows the temporal evolution of the fundamental pulse while exhibiting the expected nonlinear intensity dependence, with the second-harmonic intensity approximately proportional to the square of the fundamental intensity. This confirms efficient frequency conversion and demonstrates the pulse-shaping capability of the system for CRBS applications.

\section{Application of the portable laser system} 

\begin{figure}[h!]
    \centering
    \includegraphics[width=1\textwidth]{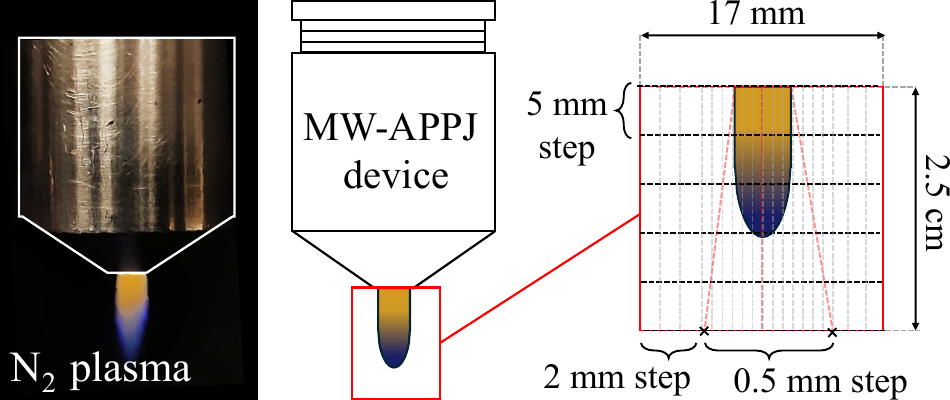}
    \caption{Image of the N$_2$ MW atmospheric plasma jet and scanning procedure for the 2D spatial map.}
    \label{fig:jet_scann}
\end{figure}

The portability of the newly developed system enables CRBS diagnostics in field environment where previous table-top systems cannot be deployed\cite{Bak2022DualAgile, 11298376}. To validate its capability of providing laboratory-grade measurements \textit{in-situ}, the system is used to characterize a commercial atmospheric-pressure microwave plasma jet\footnote{Heuermann HF-Technik, PS series} at an operating power of $75$~W and total N$_{2}$  flow of  $5$ L/min. Figure.~\ref{fig:jet_scann} shows an image of the N$_2$ plasma plume with the measurement points used for the spatial scan. Each recorded point represents an average of the number density and translational temperature over the probe volume defined by the overlap region of the pump and probe beams, which is estimated to have a cross-sectional diameter of approximately $300$~$\mu$m and a lattice length of $2$~mm.

\begin{figure}[htbp]
    \centering

    \begin{subfigure}{0.49\textwidth}
        \centering
        \includegraphics[width=1\textwidth]{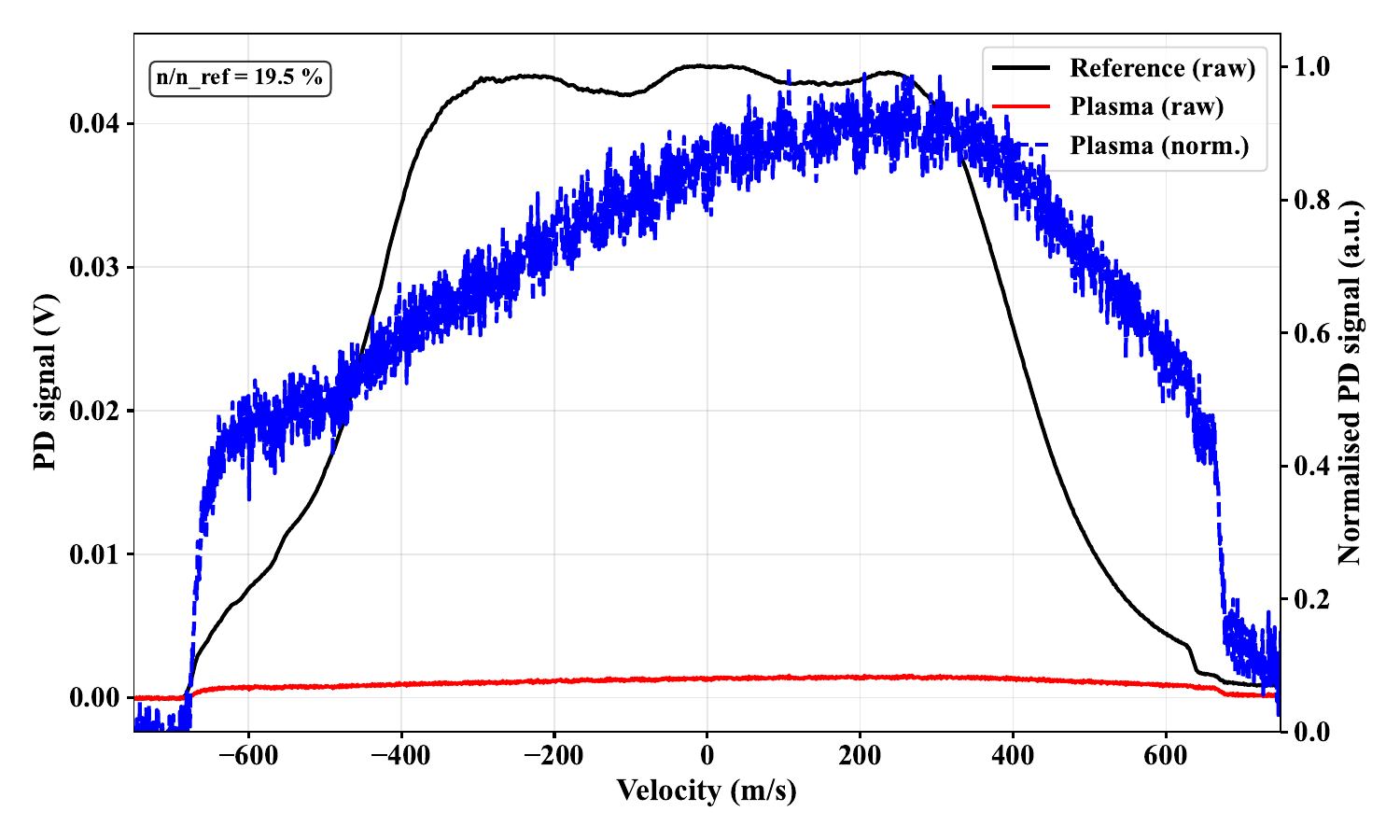}
        \caption{\label{spectrum} CRBS spectra obtained in ambient air at $293$~K, $1$~atm as reference (black), within the N$_{2}$ plasma plume at coordinates (0, 9) (red) and its amplitude normalized spectrum (blue).}
    \end{subfigure}
    \hfill
    \begin{subfigure}{0.49\textwidth}
        \centering
        \includegraphics[width=1\textwidth]{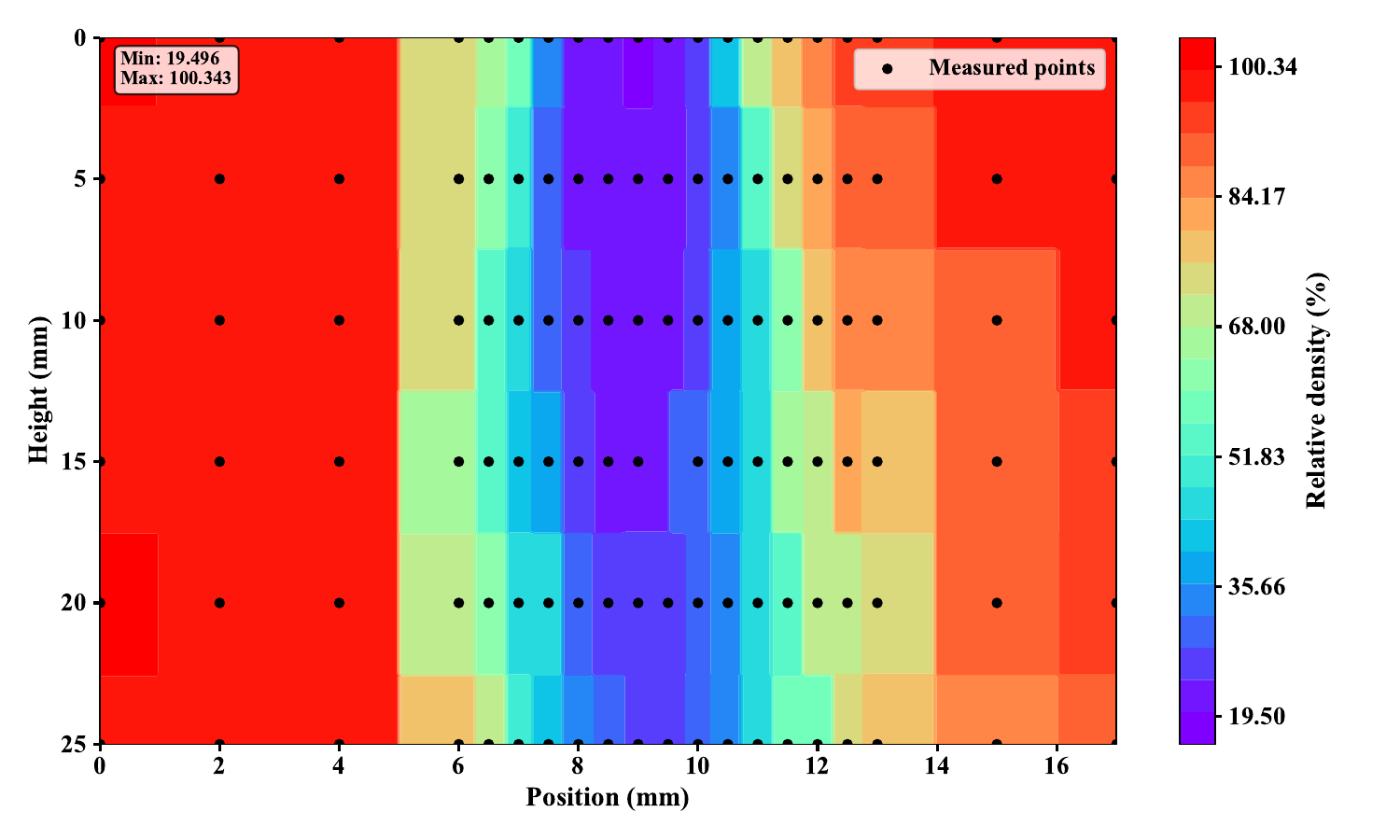}
        \caption{\label{Tfield} Spatially resolved number density map of the N$_2$ microwave plasma jet reconstructed from CRBS measurements. Black dots indicate the measurement locations.}
    \end{subfigure}
    \caption{\label{plasmajetscan}
CRBS spectra and corresponding density profile of the plasma plume.}

\end{figure}

As shown in Fig.~\ref{plasmajetscan}(a), CRBS spectra (50 averages are taken during measurements) were obtained for ambient room conditions and within the atmospheric plasma plume at position (0, 9) (see Fig.~\ref{plasmajetscan}(b)). A CRBS spectrum is obtained in room air (without plasma) serving as a calibration reference, exhibiting one Rayleigh peak on the middle with two well-defined Brillouin peaks. In contrast, the spectrum acquired from the plasma plume shows a significant decrease in peak intensity and a broadening of the spectral lineshape, where Brillouin peaks are suppressed, as expected at lower densities~\cite{grinstead2000coherent}. This broadening is due to the increased thermal velocity of the neutral species with high temperature, illustrating the system's ability to resolve thermal dynamics in a single laser shot.

The intensity of the CRBS signal $I_{\text{CRBS}}$ exhibits a quadratic dependence on the scatterer number density $N$ on isothermal conditions, i.e., $I_{\text{CRBS}} \propto N^2$, a relationship that has been experimentally verified for a several atomic and  molecular gases~\cite{Gerakis2016}. By normalizing the plasma signal integral against the air calibration reference spectral integral, the neutral number density is obtained. The stability of AOM-EOM pulse shaping ensures that the pulse shape remains consistent throughout the whole measurement campaign, so the reference spectrum is valid for every shot in the plasma. As seen in the normalized spectrum, the broadening is not fully captured as a result of high temperature, thus an underestimation of the number density is systematically presented. 

Furthermore, under the assumption of a low-flow-rate regime where the static pressure remains equilibrated with the ambient atmosphere ($P \approx 101.3~\text{kPa}$), the translational temperature $T$ was extracted using the ideal gas law, $P = N k_B T$. The derived temperatures in the plume core reached approximately $1505\pm50$K. This temperature is a rough estimation to provide qualitative insight, and detailed methods to derive the temperature and its associated uncertainties will be presented in a subsequent work.

By scanning the plasma jet through the CRBS probe volume in both the axial and transverse directions, a two-dimensional averaged number density map is reconstructed, as shown in Fig.~\ref{plasmajetscan}(b). A fine transverse step of $0.5$~mm is used in the plume core to resolve the steep radial temperature gradients, while a coarser step of $2$~mm is used in the outer regions, and an axial step of $5$~mm is used along the jet axis as shown in Fig.~\ref{fig:jet_scann}. The mean number density field shows a highly localized core near the nozzle exit where the number density drops to $19.5\%$ with a steep radial gradient as the hot plasma mixes with the cold ambient air. As the jet propagates axially away from the nozzle, the density increases, and the density profile broadens radially, consistent with the expected convection. 
At the outer edges of the scan region the CRBS spectral integral returns to the same value as the reference calibration spectrum confirming that the gas has returned to near-ambient conditions, providing a self-consistent validation of the method. Furthermore, the asymmetry observed in the density field is consistent with the visually apparent asymmetry of the plasma plume.


\section{Conclusion}
We have developed a compact, robust, stable and portable laser system that fully preserves the high-performance capabilities required for single-shot CRBS, delivering pulse energy up to $450$~mJ with arbitrary pulse shaping and a chirp rate of $22$~MHz/ns. By integrating the master-slave injection locking into a single unit and optimizing the amplification chain with a customized baseplate, we reduced the size of the system significantly while improving maintainability. The compact laser system was used to reconstruct a two-dimensional density map of the plasma jet. The mapping of a plasma jet density distribution confirms that this system is ready for deployment in large-scale engineering facilities and other environments inaccessible to conventional laboratory setups.

\section{Back matter}

\begin{backmatter}
\bmsection{Funding} This work is supported by the Luxembourg National Research Fund project $15480342$ (FRAGOLA).

\bmsection{Acknowledgment} The authors would like to thank Dr. Anatoli Morozov of Princeton University for fruitful discussions regarding the laser system architecture.

\smallskip

\bmsection{Disclosures} The authors declare no conflicts of interest.

\end{backmatter}


\bibliography{sample}



\end{document}